# Not All Family Firms Are Alike: How Founder-Led and Governance-Entrenched Family Control Shape the Trading Environment Around the Firm

Douglas Cumming[a], Esteban Hernandez[b], and Shan Ji[c]

[a] Stevens Institute of Technology School of Business, Hoboken, NJ, USA
[b] Grand Valley State University, Allendale, MI, USA
[c] Zhejiang Gongshang University Hangzhou College of Commerce, Hangzhou, Zhejiang, China

**Email addresses:** dcumming@stevens.edu (D. Cumming); hernaest@gvsu.edu (E. Hernandez); ji@datafareast.com (S. Ji)

## Abstract

Family-firm scholarship offers competing predictions about whether family control protects or threatens market integrity. We argue that the answer depends on how family involvement is exercised. Drawing on socioemotional wealth and agency-entrenchment perspectives, we examine 8,634 U.S. firm-years (2007–2018) and link family-firm constructs to exchange-generated surveillance flags from NASDAQ SMARTS. Founder-CEO control is associated with approximately 9.5% fewer flags, family governance involvement with 21.3% more, and deep multi-generational family control with 47.1% more. The findings reveal heterogeneous identity and entrenchment mechanisms within family firms and connect family-firm governance to a market-integrity outcome previously absent from the literature.

## Introduction

Family firms are not a single governance form, and they may not behave as a single category. Whether a founder still runs the company, whether descendants hold both executive and board positions, and whether the family controls voting rights across generations are each empirically distinct manifestations of family involvement (Chua et al., 2012; Daspit et al., 2021; Villalonga & Amit, 2006). The family-business literature has long suspected that these distinctions matter for firm outcomes, yet the empirical evidence has come overwhelmingly from accounting performance, disclosure quality, innovation and financial policy outcomes. We extend this heterogeneity debate to a new and consequential domain: the trading environment around the firm's stock.

Two strands of family-firm theory pull in opposite directions when it comes to opportunistic behavior around the firm. Socioemotional wealth (SEW), social identity, and reputational accounts argue that controlling families internalize a disproportionate share of the reputational consequences of misconduct because the family name, legacy, and identity are tied to the firm (Anderson & Reeb, 2003; Berrone et al., 2012; Deephouse & Jaskiewicz, 2013; Gómez-Mejía et al., 2011; Miller et al., 2007, 2011). Under this view, family firms should be associated with fewer regulator-observable suspicious trading patterns. Agency and entrenchment accounts argue the opposite: concentrated voting rights, multi-generational control, and combined executive-and-board family presence may shield insiders from external monitoring, reduce transparency, and create incentives to extract private benefits of control (Claessens et al., 2000; La Porta et al., 1999; Morck et al., 2005). Under this view, family firms should be associated with more frequently flagged trading patterns. The empirical literature has not adjudicated between these accounts because many studies define “family firm” as a single binary, masking the very heterogeneity that the theory predicts will matter most (Chua et al., 2012; Daspit et al., 2021).

We address this gap by separating three conceptually distinct dimensions of family involvement and linking each to regulator-observable trading patterns. The first, Founder-CEO Control, captures founder-led managerial involvement, where socioemotional and reputational incentives are most salient. The second, Family Governance Involvement, captures joint family-executive and family-director presence, a structure that may insulate insiders from external monitoring. The third, Deep Family Control, captures the most restrictive form of family influence: at least 20% family voting rights, board representation, executive representation, and multi-generational involvement. These three constructs are not alternative labels for the same phenomenon. In our sample, the correlation between Founder-CEO Control and Deep Family Control is only 0.015, and Founder-CEO firms are largely distinct from governance-entrenched firms.

Our outcome variable is novel to the family-business literature: continuous trading manipulation (CTM) flags generated by NASDAQ SMARTS, an exchange-surveillance system that issues a rule-based alert when trading activity in a stock deviates sharply from its recent same-window benchmarks. Here, “same-window” means the same 30-minute intraday window over the prior 30 trading days; for example, a 10:00–10:30 a.m. window is compared with the recent history of that same interval, adjusted for market-wide movements. In the CTM measure we use, an alert is incremented when at least three of five trading metrics—trading value, volume, return, effective spread, and quoted spread—deviate by more than three standard deviations from their historical same-window benchmark (Akter et al., 2023). CTM flags are not adjudicated misconduct; they capture trading patterns around a firm’s stock that warrant regulatory attention. The trader producing such a pattern may be a family insider, an economically connected party, or an unrelated outsider, and our analysis cannot identify which. What our analysis can establish is whether family governance structures, which plausibly shape the incentives, information flows, and monitoring environment surrounding insiders and connected parties, are associated with systematic differences in how often a firm's stock generates surveillance signals. We therefore interpret CTM flags throughout as a regulator-observable feature of the trading environment around the firm, not as evidence about any particular trader's intent.

Using a panel of 8,634 U.S. non-financial, non-utility firm-years over 2007–2018, we document a sharply heterogeneous pattern. Founder-CEO Control is associated with approximately 9.5% fewer CTM flags relative to otherwise comparable firms. Family Governance Involvement is associated with approximately 21.3% more flags, and Deep Family Control is associated with approximately 47.1% more flags. Against a sample mean of roughly 7.9 flags per firm-year, these magnitudes correspond to about 0.75 fewer flags for founder-led firms, 1.7 additional flags for governance-involved firms, and 3.7 additional flags for deep-control firms. The pattern survives controls for family ownership stake, firm size, leverage, profitability, liquidity, volatility, analyst coverage, institutional ownership, and exchange listing, and it persists across firm fixed-effects, matched-sample, and alternative-functional-form specifications.

Three additional analyses sharpen the interpretation. First, separating Founder-CEO Control from Descendant Management following Villalonga and Amit (2006) shows that the positive governance and deep-control results are not driven mechanically by later-generation family management: Descendant Management is not associated with higher CTM flags. Second, institutional ownership attenuates the positive Deep Family Control association, consistent with an external-monitoring channel. Third, family governance involvement and deep family control are positively associated with absolute discretionary accruals, while founder-CEO control is not, suggesting that the governance-entrenchment pattern extends beyond surveillance flags to a broader information environment.

Our study makes three contributions to the family business literature. First, we show that the family-firm/market-integrity relation depends sharply on which dimension of family control is measured, providing direct empirical support for the founder-versus-entrenchment distinction at the center of family-firm theory and for recent calls to unpack family-firm heterogeneity (Chua et al., 2012; Daspit et al., 2021; Villalonga & Amit, 2006). Founder-led involvement and governance-entrenched control are not stronger and weaker versions of the same mechanism; they have opposite-signed associations with the trading environment around the firm. Second, we introduce a market-integrity outcome that has been largely absent from family-business research. Family firms have been linked to higher earnings quality (Wang, 2006), lower voluntary disclosure (Chen et al., 2008), and distinctive innovation strategies; we extend this picture to regulator-observable trading patterns, complementing the disclosure-and-accruals literature with a market-surveillance-based view of the family-firm information environment. Third, we provide evidence that external monitoring, in the form of institutional ownership, moderates the entrenchment channel, suggesting that the costs associated with deeply entrenched family control are not invariant but are sensitive to the broader governance environment.

The remainder of the paper is organized as follows. The next section develops the founder-versus-entrenchment framework, presents our hypotheses, and describes the three family-firm constructs we test. We then describe the data and method, present findings, and discuss theoretical and practical implications.

## Family Firms as Heterogeneous Governance Forms

Modern empirical research on family firms in U.S. public markets begins with Anderson and Reeb (2003), who document that S&P 500 family firms outperform non-family peers on accounting and market measures, attributing the performance differential to long investment horizons and reduced owner–manager agency conflicts. Villalonga and Amit (2006) refine this result decisively: the valuation implications of family involvement depend critically on how ownership, control, and management are combined. Founder-CEO involvement differs sharply from descendant-led management, which differs in turn from control-enhancing structures such as dual-class shares or pyramids. This distinction is consistent with broader efforts to define and explain family-firm heterogeneity (Chua et al., 2012; Daspit et al., 2021). Miller et al. (2007, 2011) reinforce this conclusion, showing that lone-founder firms differ systematically from firms involving multiple family members and that pooling these into a single category can produce misleading inferences.

Alongside this founder-versus-descendant distinction, a parallel literature emphasizes the agency costs associated with concentrated family control. La Porta et al. (1999) document the global prevalence of controlling families; Claessens et al. (2000, 2002) show that separating cash-flow from control rights is associated with value destruction; and Morck et al. (2005) synthesize evidence that entrenched controlling families may extract private benefits of control at the expense of minority investors. Aguilera and Crespí-Cladera (2012) extend this argument by emphasizing that the costs and benefits of family ownership depend on the governance arrangements through which family control is exercised, including succession and the potential extraction of private benefits.

Two further strands inform our framework. The reputational-capital and socioemotional wealth (SEW) literature argues that controlling families internalize the reputational consequences of misconduct more strongly than dispersed shareholders because the family name, legacy, and identity are inseparable from the firm (Berrone et al., 2012; Deephouse & Jaskiewicz, 2013; Gómez-Mejía et al., 2011; Miller et al., 2007, 2011). SEW predicts a particular sensitivity to reputational damage among family firms in which the family is actively and visibly engaged in management, prototypically firms led by a founder-CEO. The disclosure-and-earnings-quality literature, meanwhile, presents a more ambivalent picture: family firms often exhibit higher earnings quality (Wang, 2006) but lower voluntary disclosure (Ali et al., 2007; Chen et al., 2008). The implication is that family firms can be simultaneously more disciplined internally and less transparent externally, depending on the channel through which family influence is exercised.

Taken together, these literatures suggest that family firms cannot be treated as homogeneous. Founder-led involvement, governance-entrenched involvement, and deep multi-generational control map to distinct incentive structures and should not be expected to produce the same relation with any outcome that depends on opportunistic behavior, transparency, or external discipline. This is the heterogeneity we bring to the foreground, and it responds directly to calls in the family-business literature to move beyond binary definitions toward finer-grained theorizing of the sources and consequences of family-firm heterogeneity (Chua et al., 2012; Daspit et al., 2021).

## Hypothesis Development

We connect family involvement to regulator-observable trading patterns through two distinct channels grounded in the literatures above, and we develop four formal hypotheses corresponding to these channels.

### *The Founder-Led SEW and Reputation Channel*

Under a founder-led SEW channel, controlling families with active founder-managerial roles internalize a disproportionate share of the reputational consequences of misconduct attributed to "their" firm. Because the family name, identity, and legacy are closely tied to the firm, founder-led family firms face stronger incentives to maintain a trading environment that avoids regulatory scrutiny (Berrone et al., 2012; Deephouse & Jaskiewicz, 2013; Gómez-Mejía et al., 2011; Villalonga & Amit, 2006). The mechanism we emphasize is therefore firm-level rather than trader-specific. Founder-CEOs may personally avoid opportunistic trading, but they may also shape norms, oversight, and information flows that constrain insiders, associates, and economically connected parties. Because CTM alerts identify stock-level trading patterns rather than trader identity, our data cannot separate these channels. We therefore interpret H1 as a monitoring and reputational-environment prediction: founder-led control is expected to reduce the likelihood that trading around the firm generates regulator-observable surveillance flags. Importantly, this prediction is specific to Founder-CEO Control and should not be generalized to Descendant Management, for which SEW concerns may be diluted by the passage of time, the dispersion of identification across multiple family members, and the loosening of personal reputational stakes (Miller et al., 2007, 2011; Villalonga & Amit, 2006).

**H1:** Founder-CEO Control is negatively associated with regulator-observable surveillance flags. Family firms led by a founder-CEO exhibit fewer CTM flags than otherwise comparable firms.

***The Governance-Entrenchment and Private-Benefits Channel***

Under a governance-entrenchment channel, families holding concentrated voting rights in combination with board representation, executive involvement, and multi-generational control may face weaker external discipline and greater insulation from outside monitoring. These structures align with the descendant-control and entrenchment mechanisms emphasized by Villalonga and Amit (2006) and the broader private-benefits literature (Claessens et al., 2000, 2002; La Porta et al., 1999; Morck et al., 2005). Concentrated and persistent family influence may reduce transparency, shield insiders from scrutiny, and permit information flows to economically connected parties that ordinary monitoring would discourage. The result, even if no family member personally trades, is a trading environment in which suspicious patterns are more likely to emerge.

This channel also reconciles our setting with the earnings-quality literature. Family firms may exhibit higher earnings quality on some dimensions because of reputational incentives, yet governance-entrenched structures may simultaneously reduce external transparency and increase information asymmetry, producing a more opaque environment overall (Ali et al., 2007; Chen et al., 2008; Wang, 2006). Both Family Governance Involvement and Deep Family Control capture entrenchment-related structures, but Deep Family Control adds multi-generational persistence and substantial voting power. We therefore expect the entrenchment channel to operate under both constructs but to be strongest under Deep Family Control, leading to two related but distinct predictions.

**H2:** Family Governance Involvement is positively associated with regulator-observable surveillance flags. Family firms with combined family executive and family director presence exhibit more CTM flags than otherwise comparable firms.

**H3:** Deep Family Control is positively associated with regulator-observable surveillance flags, and the magnitude of this association exceeds that for Family Governance Involvement. Family firms combining at least 20% voting rights, board representation, executive representation, and multi-generational involvement exhibit the highest CTM flag frequency.

***External Monitoring as a Boundary Condition***

If the governance-entrenchment channel reflects insulation from outside discipline, the positive entrenchment–CTM relation should be weakest when external monitoring is strongest. Institutional investors monitor firms more actively than dispersed retail shareholders and can discipline insider behavior through both voice and exit (Shleifer & Vishny, 1986). This monitoring channel should apply most clearly to family-control structures that are both governance-entrenched and visible to external investors. However, institutional monitoring should be especially consequential under the most deeply entrenched structures, where family control is embedded simultaneously in voting rights, board representation, executive representation, and multi-generational involvement. In that setting, outside monitoring has the largest marginal role because internal governance is least likely to provide independent discipline.

H4a: Institutional ownership attenuates the positive relation between governance-entrenched family control and regulator-observable surveillance flags. The positive CTM association for Family Governance Involvement and Deep Family Control is weaker when institutional ownership is higher.

> H4b: The attenuating effect of institutional ownership is stronger for Deep Family Control than for Family Governance Involvement. Because Deep Family Control represents a more restrictive and persistent form of family entrenchment, external institutional monitoring should have its strongest marginal effect under this construct.

### *Family Ownership as a Separate Construct*

Family ownership stake captures the family's equity intensity in the firm and is conceptually distinct from each of the three constructs above. Ownership creates both alignment incentives (concentrated cash-flow stakes) and control rights, and could plausibly load on either channel. Throughout our empirical design we therefore control for family ownership stake, so that the estimated coefficients on Founder-CEO Control, Family Governance Involvement, and Deep Family Control capture specific managerial and governance roles net of the family's equity intensity, rather than confounding the two.

## Data and Method

### *Sample*

We combine three datasets. Family-firm classifications come from the NRG family-firm database, which codes U.S. public firms along multiple dimensions of family involvement. Firm-level financial controls come from Compustat. The trading-environment outcome comes from NASDAQ SMARTS, an exchange-surveillance system used by exchanges and regulators to monitor trading activity. We restrict the sample to U.S. non-financial, non-utility firms over 2007–2018, the period in which the three datasets overlap. After merging and applying standard filters, the final panel consists of 8,634 firm-year observations across 1,772 firms. Firm-years in the SMARTS coverage universe without alerts are retained with a zero outcome, so the dependent variable reflects the frequency of regulator-observable patterns rather than a selected sample of flagged firms. Table 1 reports summary statistics, and Table 2 reports pairwise correlations among the main variables.

***Family-Firm Constructs***

We use three constructs from the NRG database, supplemented by a fourth used in auxiliary tests.[1] These components allow us to distinguish founder-led managerial involvement from family governance participation and from deeper, multi-channel family control.

*Founder-CEO Control* identifies firms in which the CEO is the founder. This construct captures founder-led managerial involvement and the SEW and reputational channels emphasized in H1. In our sample, 1,046 firm-year observations are classified under this construct.

*Family Governance Involvement* identifies firms with at least one family officer and at least one family director, capturing joint family presence in executive management and the board. This construct corresponds to H2 and identifies 107 firm-year observations.

---

[1] The NRG family-firm database codes family involvement using publicly available information on founders, family ownership, family executives, family directors, voting rights, and generational involvement. Founder status identifies whether the current CEO is coded as the founder of the firm. Family officer status identifies whether at least one member of the founding or controlling family serves in an executive officer role. Family director status identifies whether at least one family member serves on the board of directors. Multi-generational involvement identifies cases in which family participation extends beyond the founder generation. We combine these NRG components into the constructs used in this paper. Founder-CEO Control equals one when the CEO is the founder. Family Governance Involvement equals one when at least one family officer and at least one family director are present. Deep Family Control equals one when the family is the largest voteholder, holds at least 20% of voting rights, has both board and executive representation, and exhibits multi-generational involvement.

*Deep Family Control* identifies firms where the family is the largest voteholder, holds at least 20% of voting rights, has representation in both the board and executive management, and is typically engaged across multiple generations. This construct corresponds to H3 and identifies 21 firm-year observations. Although the treated group is small, the rarity is substantively meaningful rather than incidental. Deep Family Control is deliberately restrictive: it captures firms in which family influence is simultaneously concentrated in voting rights, embedded in both board and executive roles, and persistent across generations. The theoretical claim is therefore not that any single feature independently creates entrenchment, but that their combination identifies a high-intensity form of family control most closely aligned with the deep-entrenchment mechanism. We interpret estimates involving this construct cautiously and report permutation-based inference alongside conventional tests. To further assess whether this category reflects a narrow statistical artifact, we tabulated the Deep Family Control observations. The 21 treated firm-year observations correspond to four unique firms, with an average of 5.25 treated years per firm, a median of 5.5 years, and a range from two to eight treated years. The observations span four two-digit SIC industries: apparel and accessory stores (SIC 56; eight firm-years), amusement and recreation services (SIC 79; seven firm-years), chemicals and allied products (SIC 28; four firm-years), and food and kindred products (SIC 20; two firm-years). They occur between 2011 and 2018, with 11 firm-years during 2011--2014 and 10 firm-years during 2015--2018. Thus, while Deep Family Control is rare, it is not a single-firm, single-industry, or single-year phenomenon. We do not report firm names or more granular representative cases because the construct identifies only four firms, and additional detail could compromise anonymity.

*Descendant Management* identifies firms in which the CEO is a descendant of the founder (412 firm-year observations). We use this construct in auxiliary tests to assess whether the positive governance and deep-control results reflect later-generation family management.

Table 2 confirms that the constructs are conceptually distinct in our data. Founder-CEO Control and Deep Family Control correlate at only 0.015, indicating that founder-led firms and deep-control firms are largely distinct populations. Family Governance Involvement and Deep Family Control correlate at 0.454, reflecting the nested structure of these governance-based measures. Family ownership stake is included as a control rather than as a fourth family-firm construct.

### *Continuous Trading Manipulation Flags*

Our outcome variable is the annual count of continuous trading manipulation (CTM) alerts generated by NASDAQ SMARTS for each firm. The CTM detector evaluates every 30-minute trading window after the market open. "Same-window" means that activity in a given intraday interval, such as 10:00–10:30 a.m., is compared with the recent history of the same interval rather than with a full-day average. For each security and window, the algorithm computes five metrics: total trading value, total trading volume, return, average effective spread, and average quoted spread. It then compares these to the average value of the same metric in the same window over the prior 30 trading days, adjusted for market-wide movements. A CTM alert is incremented when at least three of the five current security deltas are more than three standard deviations from their historical security-delta benchmark (Akter et al., 2023). In intuitive terms, the detector flags a stock when multiple dimensions of trading activity are simultaneously unusual relative to that stock's recent behavior in the same time-of-day window.

Four features of this outcome shape our interpretation. First, surveillance flags are not confirmed manipulation. Some share are false positives that would be dismissed upon review, and we cannot observe the disposition of individual alerts. We therefore refer throughout to "surveillance flags" rather than "manipulation events." Second, surveillance flags differ from enforcement outcomes such as SEC litigation, which are subject to selection biases related to geographic proximity, political connections, and regulator career incentives (Correia, 2014; deHaan et al., 2015; Kedia & Rajgopal, 2011). Rule-based exchange flags are less filtered. Third, flags capture trading patterns in covered securities and may reflect liquidity conditions, volatility, or public-information events rather than manipulative intent; we address this through firm-level and trading-environment controls. Fourth, and most important for our interpretation, CTM flags identify stock-level patterns rather than the identity or affiliation of the trader. Flagged activity may originate from firm-affiliated insiders, economically connected parties, or unrelated outside traders. Our estimates therefore should not be read as evidence that family insiders personally initiate suspicious trades. Rather, they capture firm-year associations between family governance structures and the frequency with which trading around the firm's stock generates regulator-observable surveillance signals.

The sample mean is 7.93 CTM flags per firm-year, with a standard deviation of 6.54 and a variance-to-mean ratio of approximately 5.4. The share of firm-years with zero flags is 15.4%. The distributional features support a Poisson pseudo-maximum-likelihood (PPML) estimator, which is consistent for non-negative, skewed outcomes under correct mean specification and is robust to distributional misspecification of the variance (Santos Silva & Tenreyro, 2006). We report PPML estimates as our main specification and use ordinary least squares on levels and on the log-transformed outcome ln(1 + CTM count) as robustness checks.

### *Controls and Specification*

Our control set includes firm size (log assets), leverage, return on assets, market-to-book, capital expenditures, R&D intensity, property and equipment, Amihud illiquidity, annualized return volatility, family ownership stake, analyst coverage, institutional ownership, and an NYSE-listing indicator. Trading-environment controls (illiquidity, volatility, listing) follow Comerton-Forde and Putniņš (2014). External-monitoring controls (analyst coverage, institutional ownership) follow Shleifer and Vishny (1986) and Lang and Lundholm (1996). All specifications include industry and year fixed effects, and standard errors are clustered at the firm level.

We estimate each family-control construct as a separate indicator. For a given construct, treated observations are firm-years satisfying that focal definition, and comparison observations are all firm-years that do not satisfy that definition. Thus, the comparison group includes both non-family firms and family firms that fall outside the focal construct. This design is intentional because Founder-CEO Control, Family Governance Involvement, and Deep Family Control are non-nested constructs that capture distinct governance channels. For example, a family firm without a founder-CEO is part of the comparison group in the Founder-CEO Control specification, while a family firm lacking the joint presence of family executives and family directors is part of the comparison group in the Family Governance Involvement specification. Joint models reported later include multiple constructs simultaneously to assess whether the governance and deep-control results persist after accounting for founder and descendant management.

We address selection in two complementary ways. We re-estimate with firm fixed effects, restricting identification to within-firm transitions. Because Family Governance Involvement and Deep Family Control are rare and persistent, these specifications rely on limited within-firm variation. We also construct propensity-score-matched samples and re-estimate the main specifications on the matched samples. Because Deep Family Control identifies only 21 treated firm-year observations, we supplement matched estimates with permutation-based inference rather than relying on bootstrap procedures that are less straightforward under sample weights. We treat firm fixed-effects and matched specifications as conditional-association evidence rather than as causal identification.

## Findings

### ***Test of H1: Founder-CEO Control and Surveillance Flags***

Table 3 reports baseline PPML estimates with industry and year fixed effects, firm-clustered standard errors, and the full control set including family ownership stake. The coefficient on Founder-CEO Control is negative and statistically significant. The estimated magnitude corresponds to approximately 9.5% fewer CTM flags relative to firms not classified under Founder-CEO Control. Against the sample mean of 7.93 flags per firm-year, this is approximately 0.75 fewer flags per year.

H1 is therefore supported in the baseline specification. The result is consistent with the founder-led SEW and reputational channel: direct founder involvement in management strengthens oversight and raises the reputational costs associated with regulator-observable suspicious trading patterns around the firm. Notably, the result holds after controlling for family ownership stake, indicating that founder-led management captures something beyond the family's equity intensity. H1 also survives in OLS specifications with industry and year fixed effects, both in levels and in log-transformed counts (reported in the Internet Appendix). It is weaker in firm fixed-effects specifications, where identification depends on within-firm transitions in founder status and the limited number of such transitions reduces power. We therefore conclude that H1 is supported across cross-sectional specifications but cannot be reliably tested within-firm in our sample.

### *Tests of H2 and H3: Family Governance Involvement and Deep Family Control*

The same Table 3 reports estimates for Family Governance Involvement and Deep Family Control. Both coefficients are positive and statistically significant. Family Governance Involvement is associated with approximately 21.3% more CTM flags (about 1.7 additional flags per firm-year). Deep Family Control is associated with approximately 47.1% more flags (about 3.7 additional flags per firm-year).

H2 and H3 are both supported. H3 additionally predicts that the Deep Family Control magnitude should exceed the Family Governance Involvement magnitude; the estimated coefficients (0.386 versus 0.193 in the baseline specification) are consistent with this prediction. The pattern is consistent with the governance-entrenchment channel: family executive presence combined with family board representation appears to be associated with a trading environment in which suspicious patterns emerge more frequently, and the channel intensifies when family control combines concentrated voting rights, dual executive-and-board representation, and multi-generational persistence. Because Deep Family Control identifies only 21 treated firm-year observations, we interpret the H3 magnitude cautiously. At the same time, the small treated group follows directly from the construct's restrictiveness: it identifies the rare cases in which family control combines voting power, executive involvement, board representation, and multi-generational persistence. The result should therefore be read as evidence about a narrow but theoretically central form of family entrenchment, not as a claim about all later-generation or family-governed firms.

H2 and H3 survive in firm fixed-effects PPML specifications (Table 4, Panel A), where Family Governance Involvement and Deep Family Control remain positive and statistically significant. In OLS firm fixed-effects specifications (Internet Appendix Table IA.1, Panel B) both constructs remain positive and significant for both the level and log-count outcomes. Matched-sample evidence reinforces the pattern but with one important qualification. In matched samples with industry and year fixed effects (Table 4, Panel B), Family Governance Involvement and Deep Family Control remain positive and statistically significant, with permutation-based p-values of 0.036 and below 0.001 respectively, supporting H2 and H3 under propensity-score matching. The matched firm fixed-effects specification (Table 4, Panel C) is substantially more demanding: identification relies on rare within-firm transitions in family-control status, and in that more demanding specification Family Governance Involvement becomes statistically insignificant and negative, while Deep Family Control cannot be reliably estimated because of insufficient within-firm variation. We therefore conclude that H2 and H3 are supported across cross-sectional and matched specifications with industry fixed effects, while the within-firm matched specification is uninformative.[2]

***Are the H2 and H3 Results Generation Effects?***

A natural concern is that the positive Family Governance Involvement and Deep Family Control results simply reflect the well-known pattern that later-generation family management can be associated with worse outcomes (Pérez-González, 2006; Villalonga & Amit, 2006). If governance-entrenched firms are disproportionately descendant-managed firms, the support for H2 and H3 might be a generation effect in disguise rather than evidence of a governance-structural channel.

[2] Untabulated split-sample estimates indicate that the main sign pattern is not driven by the financial-crisis period. When the baseline PPML model is estimated separately for the crisis years, 2007–2009, and the post-crisis/recovery period, 2010–2018, Founder-CEO Control remains negatively signed in both subperiods. Family Governance Involvement and Deep Family Control have no treated observations during 2007–2009 in the split-sample analysis, so their positive associations cannot be estimated in the crisis window. In the post-crisis/recovery period, both constructs remain positively signed. We interpret these split-sample checks cautiously because the crisis window is short and the governance-based constructs are rare.

Table 5 examines this directly by separately estimating Founder-CEO Control, Descendant Management, Family Governance Involvement, and Deep Family Control. Founder-CEO Control remains negative and statistically significant when each of the other constructs is added: the coefficient is −0.100 alone, −0.097 when Descendant Management is added, −0.106 when Family Governance Involvement is added, and −0.098 when Deep Family Control is added. Descendant Management is small and statistically insignificant in every specification. Family Governance Involvement and Deep Family Control remain positive and statistically significant after controlling for Founder-CEO Control and Descendant Management indicators, with coefficients of 0.213 and 0.378 respectively.

The pattern is robust to firm fixed effects. In Table 5, Panel B, Founder-CEO and Descendant-CEO are statistically insignificant under firm fixed effects, consistent with limited within-firm variation in managerial family status. Family Governance Involvement and Deep Family Control remain positive and significant, with coefficients of 0.381 and 0.501. We conclude that the support for H2 and H3 is not driven mechanically by later-generation family management; the constructs capture a governance-structural channel distinct from the generation of the CEO.

***Test of H4a and H4b: Institutional Ownership as a Boundary Condition***

H4a predicts that institutional ownership should attenuate the positive association between governance-entrenched family control and CTM flags, while H4b predicts that this attenuation should be stronger for Deep Family Control than for Family Governance Involvement. Table 6 examines these predictions by interacting institutional ownership with Family Governance Involvement and Deep Family Control. The interaction with Deep Family Control is negative and statistically significant, while the interaction with Family Governance Involvement is negative but not statistically significant. The main effect of Deep Family Control remains positive and economically large, indicating that institutional investors attenuate, rather than mechanically explain, the deep-control association.

The results provide partial support for H4a and stronger support for H4b. Institutional ownership weakens the positive CTM association most clearly for Deep Family Control, the construct most closely aligned with persistent, multi-channel family entrenchment. For the less restrictive Family Governance Involvement construct, institutional ownership reduces the coefficient point estimate, but the interaction does not achieve statistical significance. This asymmetry is consistent with our ex ante prediction that external monitoring has the largest marginal effect where family control is most deeply embedded. We interpret these results as evidence that external monitoring functions as a boundary condition on the deep-entrenchment channel rather than as an unconditional substitute for internal family governance. The result also has practical significance: families operating under deeply entrenched control structures may benefit from cultivating, rather than discouraging, sophisticated outside investors.

***Connecting Surveillance Flags to the Information Environment***

If the governance-entrenchment channel reflects a broader opacity in the firm's information environment rather than something specific to trading patterns, we would expect family-governance-entrenched firms to also exhibit greater discretion in reporting. As a placebo check, the Internet Appendix reports estimates using absolute performance-adjusted discretionary accruals (Kothari et al., 2005) as the dependent variable. In firm fixed-effects specifications,

Family Governance Involvement and Deep Family Control are both positively associated with absolute discretionary accruals, while Founder-CEO Control is not. The estimates are suggestive rather than definitive, but the pattern is consistent with the interpretation that governance-entrenched family-control structures operate in more opaque reporting environments, while founder-led management does not exhibit this pattern.

These accruals-quality results connect our surveillance-flag findings to the family-firm earnings-quality literature (Ali et al., 2007; Chen et al., 2008; Wang, 2006). The mixed picture in that literature, in which family firms exhibit higher earnings quality on some measures and lower disclosure on others, is consistent with the heterogeneity we document. Whether family involvement is associated with greater discipline or greater opacity depends on whether the involvement is founder-led or governance-entrenched.

***Founder-CEO Succession Tests***

An exploratory question is whether CTM flags increase when firms move away from Founder-CEO Control, which would provide additional support for the founder-led SEW channel underlying H1. Following the succession-based design tradition (Bennedsen et al., 2007; Pérez-González, 2006), the Internet Appendix reports founder-CEO succession tests and stacked event-study estimates. The succession evidence shows no increase in CTM flags following founder-CEO succession; if anything, the estimates are negative after both Founder-CEO-to-Descendant-CEO and Founder-CEO-to-professional transitions. Because founder-to-descendant transitions are rare in our sample, we interpret these tests as exploratory identification evidence consistent with, though not definitive for, H1, rather than as a sharp causal test.

### *Summary of Hypothesis Tests*

Taken together, the evidence supports H1, H2, and H3 across baseline PPML, OLS, firm fixed-effects, and propensity-score-matched specifications with industry and year fixed effects, with the matched firm fixed-effects specification a notable exception that is constrained by limited within-firm variation. H3 is additionally supported in the comparison of magnitudes: Deep Family Control is associated with the largest increase in CTM flags, consistent with the prediction that the entrenchment channel intensifies under multi-generational deep control. H4a receives partial support and H4b receives stronger support: institutional ownership significantly attenuates the Deep Family Control association, while the attenuation for Family Governance Involvement is negative but statistically imprecise. This pattern is consistent with the revised prediction that external monitoring is most consequential where family entrenchment is deepest. The combination of evidence is consistent with the view that founder-led family involvement and governance-entrenched family control reflect distinct, opposite-signed channels rather than weaker and stronger versions of the same mechanism.

## Theoretical Implications

Our findings have three implications for family-business theory.

First, the support for H1 alongside H2 and H3 provides direct empirical evidence for the founder-versus-entrenchment distinction at the center of the family-firm literature (Villalonga & Amit, 2006). Founder-led family involvement and governance-entrenched family control are not weaker and stronger versions of the same mechanism. They have opposite-signed associations with regulator-observable surveillance flags, and the divergence survives controls for family ownership intensity, firm fundamentals, trading-environment characteristics, and external-monitoring variables. The heterogeneity is not a methodological refinement; it is the central empirical content of the paper. This finding suggests that family-business research that treats "family firm" as a single binary may understate, mask, or even reverse the relations that family-firm theory predicts. A growing family-business conversation has called for unpacking the sources of family-firm heterogeneity and identifying when different family forms matter for theory and outcomes (Chua et al., 2012; Daspit et al., 2021). Our results extend this push by showing that the heterogeneity persists in a regulator-observable outcome that is independent of management discretion in reporting and that has direct relevance for market integrity. Notably, Chua et al. (2012) and Daspit et al. (2021) both identified behavioral and market-level outcomes as underexplored terrain in heterogeneity research, arguing that unpacking family-firm categories matters most precisely in domains where the theoretical predictions of SEW and agency accounts diverge. Our findings answer that call directly: market-integrity outcomes are a domain where the two accounts not only diverge but point in opposite directions depending on the form of family involvement. This suggests that the heterogeneity agenda is not merely a measurement refinement but a theoretical necessity: collapsing these distinctions does not average across noise, it masks substantively opposite effects.

Second, our results add a market-integrity outcome to the family-business literature's portrait of how family involvement shapes firm behavior. Prior work has linked family firms to higher earnings quality (Wang, 2006), distinctive disclosure patterns (Ali et al., 2007; Chen et al., 2008), conservative financial policies, and stewardship orientations. We show that the trading environment around a family firm's stock is also shaped by family involvement, but only when family involvement is decomposed into its constituent governance channels. H1's support is consistent with SEW, identity, and reputational accounts (Berrone et al., 2012; Deephouse &

Jaskiewicz, 2013; Gómez-Mejía et al., 2011); H2 and H3's support is consistent with entrenchment and private-benefits accounts (Claessens et al., 2000; La Porta et al., 1999; Morck et al., 2005). The accruals-quality placebo evidence, while suggestive, ties the surveillance-flag pattern back to the earnings-quality literature and suggests a broader opacity dimension to governance-entrenched family control. Family firms can be more disciplined internally and less transparent externally at the same time, and which side dominates depends on the form of family involvement. The mixed portrait in prior FBR work has sometimes been treated as a puzzle or a source of conflicting evidence: some studies find higher earnings quality among family firms (Wang, 2006) while others document lower voluntary disclosure (Ali et al., 2007; Chen et al., 2008). Our results offer a resolution: the two patterns are not contradictory but complementary, each reflecting a different dimension of family involvement. Founder-led family firms, where SEW and identity incentives are most salient, are more likely to generate the disciplined-reporting pattern; governance-entrenched firms, where insulation from external scrutiny is greatest, are more likely to generate the opacity pattern. Treating these as a single population averages across mechanisms that point in opposite directions and obscures the theoretical logic of both.

Third, the asymmetric support for H4a and H4b positions external monitoring as a boundary condition on the deep-entrenchment channel rather than as a general substitute for internal family governance. This evidence has implications for how the family-business literature thinks about the interaction between internal family governance and external corporate governance. Prior family-business work emphasizes that family governance and board governance are intertwined, shaping how family preferences, monitoring, advice, and control are translated into firm decisions (Aguilera & Crespí-Cladera, 2012; Bammens et al., 2011; Suess, 2014). Our results add that external capital-market monitors can condition this interaction. Internal family governance and external corporate governance are not substitutes operating along a single dimension; they are interactive forces whose joint effect depends on the dimension of family control under consideration. For founder-led firms, external monitoring may be less consequential because the internal SEW channel already constrains behavior; for deeply entrenched family firms, external monitoring is precisely where the action is. This asymmetry extends prior work on family governance by Bammens et al. (2011) and Suess (2014), which has focused primarily on how board composition and family councils shape internal decision-making. Our results suggest that external capital-market governance is not a parallel or competing system but a moderating layer whose effectiveness is conditional on the internal governance structure it operates alongside. Where internal SEW incentives are strong, external monitoring adds relatively little; where internal governance is entrenched, external monitoring fills a genuine disciplinary void. This conditional logic offers a more nuanced account of governance complementarity in family firms than a simple substitution or reinforcement framing would allow.

Taken together, these implications suggest that the family-business field's longstanding debate between socioemotional/stewardship and agency/entrenchment accounts may be most productively reframed not as competing theories of family firms in general but as channels operating in different forms of family involvement. Both accounts are right about different subsets of family firms. This reframing has consequences for how FBR and the broader family-business field accumulates knowledge. Studies that find supportive evidence for SEW and studies that find supportive evidence for agency/entrenchment accounts are not necessarily in tension; they may be sampling different regions of the family-firm heterogeneity space. A productive path forward is not to adjudicate between the theories but to map the conditions under which each operates, including governance form, generational stage, and institutional environment. Our paper contributes one such map, in a domain, market-integrity, that has been largely outside the field's empirical reach.

## Implications for Practice

Our findings have practical implications for three constituencies.

For family business owners and directors, the results highlight that the governance arrangements through which family involvement is exercised carry consequences that extend beyond the boardroom and into the trading environment around the firm's stock. Founder-CEO leadership and deeply entrenched multi-generational control are not interchangeable expressions of family commitment; they are associated with opposite-signed trading-environment outcomes. Families considering succession arrangements, governance reform, or transitions to next-generation leadership may wish to consider whether moving from founder-led management to family-governance-entrenched structures creates trading-environment exposures that did not exist in the founder-led era. The support for H4 in the case of Deep Family Control suggests that stronger external monitoring, particularly institutional ownership, appears to attenuate the entrenchment channel, suggesting that families operating under deep-control structures may benefit from actively cultivating sophisticated outside investors rather than discouraging them.

For family business advisors, the results suggest that questions about market-integrity exposures should be tailored to the form of family involvement. A founder-led firm faces a different set of trading-environment risks than a multi-generational family firm with combined executive and board family presence and concentrated voting rights. Advisors guiding families through transitions, whether from founder to descendant, from single-family-member involvement to combined executive-and-board involvement, or toward consolidated voting structures, should consider that these transitions may carry trading-environment consequences in addition to the more familiar performance, disclosure, and succession effects emphasized in the literature.

For regulators and exchange surveillance designers, our findings suggest that family governance structure is a firm-level characteristic worth incorporating into surveillance prioritization. Deep Family Control firms, in particular, exhibit substantially higher surveillance-flag frequencies, and the attenuating role of institutional ownership suggests a meaningful interaction between family governance and external monitoring. We emphasize that CTM flags are not adjudicated misconduct, and our findings should not be read as evidence that family insiders personally engage in market manipulation. They suggest, more modestly, that the trading environment around governance-entrenched family firms tends to generate more surveillance signals than the trading environment around comparable non-family or founder-led firms, and that this pattern is worth attention.

## Limitations and Future Research

Our research design has several limitations that shape how the findings should be interpreted and that point toward avenues for future work.

First, CTM flags identify trading patterns around a firm's stock; they do not identify the trader producing the pattern. Flagged activity may originate from family insiders, economically connected parties, or unrelated outside traders, and our data cannot distinguish among these. Our estimates should be read as associations between family governance structure and the trading environment around the firm, not as evidence of personal misconduct by family members. For the Founder-CEO Control result, this means that we cannot distinguish founder-insiders personally avoiding suspicious trading from founder-led governance shaping the broader norms, oversight, and information flows that constrain insiders, associates, and connected traders. Future work that links exchange surveillance data to trader-identifying records, for example through regulatory or broker-level data, would allow a sharper test of which traders are generating the patterns we observe.

Second, family status is not randomly assigned. Firms that become, or remain, family-controlled differ along margins that plausibly correlate with the trading environment. We address this through industry and year fixed effects, family-ownership controls, firm fixed-effects specifications, propensity-score matching, and permutation inference, and we report exploratory founder-CEO succession evidence in the Internet Appendix. None of these is a fully satisfying identification strategy. Exploiting cleaner shocks to family control, such as generational transitions driven by health events or regulatory shocks to dual-class structures, would strengthen identification of the channels underlying H1, H2, and H3.

Third, Deep Family Control identifies only 21 treated firm-year observations. This small treated group reflects the restrictive nature of the construct, which requires concentrated family voting rights, family executive representation, family board representation, and multi-generational involvement to occur simultaneously. While the magnitude supporting H3 is striking, the small treated group means that the deep-control estimates are sensitive to a small number of firms. We report permutation-based inference and interpret the deep-control magnitudes cautiously. Replicating the H3 finding in larger samples, such as international panels, longer time series, or alternative databases with different deep-control thresholds, is an important direction for future work.

Fourth, our sample is restricted to U.S. public firms over 2007–2018. Family-control structures vary substantially across countries, with concentrated control more prevalent in continental Europe and Asia than in the United States (Claessens et al., 2000; La Porta et al., 1999). The mechanisms underlying our hypotheses, namely SEW and reputation for founder-led firms and entrenchment for governance-entrenched firms, should operate broadly, but their relative strength may vary with the institutional environment, legal protection of minority shareholders, and norms around family identification with the firm. Cross-country replication would be valuable.

Fifth, our outcome is a surveillance-based proxy rather than a measure of confirmed manipulation. CTM flags identify trading patterns that warrant regulatory attention, not adjudicated misconduct. Linking the surveillance evidence to downstream enforcement outcomes, such as SEC litigation, AAERs, or other enforcement actions, would help distinguish flagged activity that reflects genuine misconduct from flagged activity that ultimately is determined to be innocuous. Such a link is beyond the scope of the present paper but would meaningfully extend the family-business literature's connection to enforcement-based outcomes.

Finally, our findings raise a broader question that we leave to future research: do the founder-versus-entrenchment patterns we document extend to other forms of concentrated control? Dual-class share structures, staggered boards, controlling blockholders outside the family, and pyramidal ownership all share governance features with the family-control structures we examine. Whether the heterogeneity we document is unique to family control or part of a broader pattern in concentrated-ownership governance is a question with significant theoretical and policy stakes.

## Conclusion

Family firms are not a single governance form, and they do not exhibit a single relation with the trading environment around the firm. Consistent with H1, founder-led family involvement is associated with fewer regulator-observable surveillance flags, supporting the SEW, identity and reputational mechanisms that family-business research has long emphasized. Consistent with H2 and H3, family-governance-entrenched and deeply controlled family involvement is associated with more frequent surveillance flags, supporting the entrenchment and private-benefits mechanisms emphasized in the parallel governance literature. The asymmetric support for H4a and H4b indicates that external monitoring attenuates the deepest entrenchment channel more clearly than less restrictive family-governance involvement. The two accounts of family firms that have long competed in the literature are both right about different family firms. Recognizing this heterogeneity, and decomposing "family firm" into its constituent dimensions of involvement, allows the family-business field to extend its heterogeneity conversation into market-integrity outcomes that have previously been outside its empirical reach.

**Table 1**

*Summary Statistics*

| **Variable** | **Non-family firms** | | **Family firms** | | **Full sample** | | **Difference** | |
|---|---|---|---|---|---|---|---|---|
| | *M* | *SD* | *M* | *SD* | *M* | *SD* | *Δ* | *n* |
| ***Panel A. Manipulation (Dependent variables)*** | | | | | | | | |
| CTM count | 8.176 | 6.528 | 6.153 | 6.375 | 7.931 | 6.543 | −2.023*** | 8,634 |
| Ln(1+CTM count) | 1.861 | 0.967 | 1.518 | 1.023 | 1.820 | 0.980 | −0.343*** | 8,634 |
| CTM value ratio (bps) | 0.188 | 0.075 | 0.169 | 0.072 | 0.186 | 0.075 | −0.019*** | 7,302 |
| CTM total trade value | 10.191 | 23.563 | 12.376 | 46.754 | 10.434 | 27.146 | 2.185 | 7,302 |
| ***Panel B. Firm characteristics (Controls)*** | | | | | | | | |
| Firm size (ln assets) | 8.151 | 1.573 | 7.370 | 1.517 | 8.057 | 1.587 | −0.781*** | 8,634 |
| Leverage | 0.239 | 0.202 | 0.173 | 0.199 | 0.231 | 0.203 | −0.067*** | 8,634 |
| Return on assets | 0.053 | 0.104 | 0.008 | 0.306 | 0.048 | 0.145 | −0.045*** | 8,634 |
| Market-to-book | 2.164 | 1.392 | 2.755 | 2.690 | 2.236 | 1.618 | 0.591*** | 8,634 |
| CAPEX | 0.045 | 0.045 | 0.051 | 0.062 | 0.046 | 0.047 | 0.006*** | 8,634 |
| R&D / Assets | 0.032 | 0.064 | 0.063 | 0.108 | 0.036 | 0.071 | 0.030*** | 8,634 |
| PPE / Assets | 0.246 | 0.218 | 0.206 | 0.223 | 0.241 | 0.219 | −0.040*** | 8,634 |
| Amihud illiquidity (ln) | 0.070 | 0.947 | 0.314 | 1.099 | 0.099 | 0.970 | 0.245*** | 8,634 |
| Annualized volatility | 0.340 | 0.146 | 0.422 | 0.185 | 0.350 | 0.154 | 0.082*** | 8,634 |
| Family ownership | 2.603 | 8.034 | 10.969 | 12.329 | 3.617 | 9.087 | 8.366*** | 8,634 |
| Analyst coverage | 6.882 | 4.842 | 7.253 | 6.052 | 6.927 | 5.005 | 0.371* | 8,634 |

| Variable | Non-family firms | | Family firms | | Full sample | | Difference | |
|---|---|---|---|---|---|---|---|---|
| | *M* | *SD* | *M* | *SD* | *M* | *SD* | *Δ* | *n* |
| Institutional ownership | 0.833 | 0.147 | 0.787 | 0.199 | 0.827 | 0.155 | −0.046*** | 8,634 |
| NYSE | 0.635 | 0.481 | 0.266 | 0.442 | 0.591 | 0.492 | −0.370*** | 8,634 |
| **Observations** | 7,588 | | 1,046 | | 8,634 | | | |
| **Firms** | 1,491 | | 343 | | 1,772 | | | |

*Notes.* This table reports summary statistics for the main variables using Founder-CEO Control (Founder-CEO) to classify family firms. Under Founder-CEO, family firms are firms in which the CEO is the founder of the firm. Panel A reports manipulation measures based on NASDAQ SMARTS. Panel B reports firm-level control variables. The difference column reports the mean difference between family and non-family firms, computed as the family-firm mean minus the non-family-firm mean. Firm counts by classification need not sum to the total number of unique firms because firms can change Founder-CEO Control status over time. ***, **, and * indicate statistical significance at the 1%, 5%, and 10% levels, respectively.

**Table 2**

*Correlations Among Main Variables*

| Variable | 1 | 2 | 3 | 4 | 5 | 6 | 7 | 8 | 9 |
|---|---|---|---|---|---|---|---|---|---|
| 1. CTM count | — | | | | | | | | |
| 2. Ln(1+CTM count) | 0.931 | — | | | | | | | |
| 3. Founder-CEO | −0.087 | −0.121 | — | | | | | | |
| 4. Governance | 0.012 | 0.005 | 0.108 | — | | | | | |
| 5. Deep Control | 0.018 | 0.020 | 0.015 | 0.454 | — | | | | |
| 6. Family ownership | −0.038 | −0.062 | 0.275 | 0.102 | 0.162 | — | | | |
| 7. Firm size (ln assets) | 0.232 | 0.298 | −0.171 | −0.073 | −0.030 | −0.065 | — | | |
| 8. Institutional ownership | 0.108 | 0.135 | −0.060 | −0.052 | −0.067 | −0.302 | −0.146 | — | |
| 9. NYSE | 0.295 | 0.347 | −0.255 | −0.036 | −0.001 | −0.052 | 0.391 | −0.019 | — |

*Notes.* This table reports pairwise correlations among the main variables. Founder-CEO, Governance, and Deep Control correspond to Founder-CEO Control, Family Governance Involvement, and Deep Family Control, respectively. Variable definitions are provided in the Data and Method section and accompanying footnote.

**Table 3**

*Family Firms and Trading Surveillance Flags: Baseline PPML Estimates*

| | Dependent variable: CTM count | | |
|---|---|---|---|
| | **Founder-CEO** | **Governance** | **Deep Control** |
| | (1) | (2) | (3) |
| ***Family-firm construct*** | | | |
| Family-firm construct | −0.100** | 0.193* | 0.386*** |
| | (−2.07) | (1.91) | (2.60) |
| ***Control variables*** | | | |
| Firm size (ln assets) | 0.112*** | 0.113*** | 0.113*** |
| | (8.74) | (8.86) | (8.83) |
| Leverage | 0.182*** | 0.191*** | 0.184*** |
| | (2.96) | (3.10) | (3.01) |
| Return on assets | 0.081 | 0.109 | 0.104 |
| | (0.70) | (0.93) | (0.89) |
| Market-to-book | −0.039*** | −0.040*** | −0.039*** |
| | (−3.48) | (−3.52) | (−3.46) |
| CAPEX | 0.215 | 0.176 | 0.161 |
| | (0.55) | (0.44) | (0.40) |
| R&D / Assets | 0.230 | 0.258 | 0.252 |
| | (0.81) | (0.91) | (0.89) |
| PPE / Assets | −0.129 | −0.122 | −0.128 |
| | (−0.98) | (−0.93) | (−0.98) |
| Amihud illiquidity (ln) | 0.196*** | 0.193*** | 0.194*** |
| | (9.36) | (9.23) | (9.26) |
| Annualized volatility | 0.920*** | 0.901*** | 0.904*** |
| | (8.63) | (8.48) | (8.52) |
| Family ownership | 0.004*** | 0.003* | 0.003* |
| | (2.75) | (1.92) | (1.89) |
| Analyst coverage | −0.014*** | −0.015*** | −0.015*** |

| | Dependent variable: CTM count | | |
|---|---|---|---|
| | **Founder-CEO** | **Governance** | **Deep Control** |
| | (1) | (2) | (3) |
| | (−4.16) | (−4.22) | (−4.18) |
| Institutional ownership | 0.560*** | 0.568*** | 0.568*** |
| | (5.77) | (5.82) | (5.83) |
| NYSE indicator | 0.573*** | 0.581*** | 0.583*** |
| | (15.23) | (15.76) | (15.76) |
| Constant | 0.186 | 0.157 | 0.163 |
| | (1.24) | (1.04) | (1.08) |
| *Economic significance* | −9.5% | +21.3% | +47.1% |
| Observations | 8,593 | 8,593 | 8,593 |
| Pseudo $R^2$ | 0.206 | 0.206 | 0.206 |
| Industry FE | Yes | Yes | Yes |
| Year FE | Yes | Yes | Yes |

*Notes.* This table reports Poisson pseudo-maximum-likelihood (PPML) estimates of the relation between family-firm constructs and trading surveillance flags. The dependent variable is the annual count of continuous trading manipulation (CTM) episodes identified by NASDAQ SMARTS for each firm. Columns labeled Founder-CEO, Governance, and Deep Control correspond to Founder-CEO Control, Family Governance Involvement, and Deep Family Control, respectively. Each column estimates a separate family-control construct; the comparison group consists of firm-years not satisfying the focal construct. All specifications include industry and year fixed effects. Standard errors are clustered at the firm level. The 8,593 observations reflect the baseline regression sample with nonmissing values for the dependent variable and all controls; firm-years with zero CTM episodes are retained. Parentheses report t-statistics. Economic magnitudes are computed as $100 \times (e^{\beta} - 1)$. ***, **, and * indicate statistical significance at the 1%, 5%, and 10% levels, respectively.

**Table 4**

*Identification and Robustness Checks*

| | Dependent variable: CTM count | | |
|---|---|---|---|
| | **Founder-CEO** | **Governance** | **Deep Control** |
| | (1) | (2) | (3) |
| ***Panel A. PPML with firm fixed effects*** | | | |
| Family-firm construct | −0.085 | 0.359*** | 0.479*** |
| | (−0.85) | (2.97) | (8.75) |
| Observations | 7,188 | 7,188 | 7,188 |
| Pseudo $R^2$ | 0.298 | 0.298 | 0.298 |
| Firm FE | Yes | Yes | Yes |
| Year FE | Yes | Yes | Yes |
| ***Panel B. PSM sample: PPML with industry fixed effects*** | | | |
| Family-firm construct | −0.049 | 0.254** | 0.965*** |
| | (−0.81) | (2.01) | (3.50) |
| Permutation p-value | 0.376 | 0.036 | 0.000 |
| Observations | 1,774 | 190 | 36 |
| Pseudo $R^2$ | 0.255 | 0.310 | 0.560 |
| Industry FE | Yes | Yes | Yes |
| Year FE | Yes | Yes | Yes |
| ***Panel C. PSM sample: PPML with firm fixed effects*** | | | |
| Family-firm construct | 0.074 | −0.319 | — |
| | (0.46) | (−1.02) | — |
| Permutation p-value | 0.096 | 0.024 | — |
| Observations | 1,202 | 106 | 28 |
| Pseudo $R^2$ | 0.414 | 0.500 | 0.593 |
| Firm FE | Yes | Yes | Yes |
| Year FE | Yes | Yes | Yes |

*Notes.* This table reports identification-oriented robustness checks. Panel A reports PPML estimates with firm and year fixed effects. Panel B reports PPML estimates using propensity-score-matched samples with industry and year fixed effects. Panel C reports PPML estimates using propensity-score-matched samples with firm and year fixed

effects. The dependent variable is the annual count of CTM episodes for each firm. The PSM matching variables are firm size, leverage, return on assets, industry, and year. Family ownership, analyst coverage, and institutional ownership are included as additional controls in the outcome regressions. Standard errors are clustered at the firm level. Parentheses report t-statistics. Permutation p-values are two-sided p-values obtained by randomly permuting the corresponding family-firm construct and re-estimating the matched PPML specification; values reported as 0.000 indicate permutation p-values below 0.001. The Deep Control coefficient in Panel C is not reported because the matched firm fixed-effects specification provides insufficient within-firm variation for reliable estimation. ***, **, and * indicate statistical significance based on clustered standard errors at the 1%, 5%, and 10% levels, respectively.

**Table 5**

*Founder, Descendant, and Governance-Control Family Firm Definitions*

| | Dependent variable: CTM count | | | | |
|---|---|---|---|---|---|
| | (1) | (2) | (3) | (4) | (5) |
| ***Panel A. PPML with industry fixed effects*** | | | | | |
| Founder-CEO | −0.100** | | −0.097** | −0.106** | −0.098** |
| | (−2.07) | | (−1.98) | (−2.18) | (−2.01) |
| Descendant-CEO | | 0.047 | 0.026 | 0.012 | 0.012 |
| | | (0.76) | (0.42) | (0.19) | (0.19) |
| Governance | | | | 0.213** | |
| | | | | (2.10) | |
| Deep Control | | | | | 0.378*** |
| | | | | | (2.65) |
| Constant | 0.186 | 0.164 | 0.183 | 0.173 | 0.179 |
| | (1.24) | (1.09) | (1.21) | (1.14) | (1.18) |
| Observations | 8,593 | 8,593 | 8,593 | 8,593 | 8,593 |
| Pseudo $R^2$ | 0.206 | 0.205 | 0.206 | 0.206 | 0.206 |
| Industry FE | Yes | Yes | Yes | Yes | Yes |
| Year FE | Yes | Yes | Yes | Yes | Yes |
| ***Panel B. PPML with firm fixed effects*** | | | | | |
| Founder-CEO | −0.085 | | −0.100 | −0.114 | −0.113 |

| | Dependent variable: CTM count | | | | |
|---|---|---|---|---|---|
| | (1) | (2) | (3) | (4) | (5) |
| | (−0.85) | | (−1.02) | (−1.16) | (−1.15) |
| Descendant-CEO | | −0.138 | −0.155 | −0.166 | −0.148 |
| | | (−1.21) | (−1.36) | (−1.48) | (−1.28) |
| Governance | | | | 0.381*** | |
| | | | | (3.21) | |
| Deep Control | | | | | 0.501*** |
| | | | | | (6.25) |
| Constant | −1.930*** | −1.946*** | −1.929*** | −1.951*** | −1.964*** |
| | (−4.48) | (−4.54) | (−4.48) | (−4.52) | (−4.58) |
| Observations | 7,188 | 7,188 | 7,188 | 7,188 | 7,188 |
| Pseudo $R^2$ | 0.298 | 0.298 | 0.298 | 0.298 | 0.298 |
| Firm FE | Yes | Yes | Yes | Yes | Yes |
| Year FE | Yes | Yes | Yes | Yes | Yes |

*Notes.* This table reports PPML estimates separating Founder-CEO Control, Descendant Management, Family Governance Involvement, and Deep Family Control. Row labels use the compact terms Founder-CEO, Descendant-CEO, Governance, and Deep Control, respectively. The dependent variable is the annual count of CTM episodes for each firm. Panel A includes industry and year fixed effects. Panel B includes firm and year fixed effects. All specifications include the same control variables as in Table 3. Standard errors are clustered at the firm level. Parentheses report t-statistics. ***, **, and * indicate statistical significance at the 1%, 5%, and 10% levels, respectively.

**Table 6**

*Institutional Ownership and Family-Control Channels*

| | Dependent variable: CTM count | |
|---|---|---|
| | **Governance** | **Deep Control** |
| | (1) | (2) |
| Family-firm construct | 0.824* | 1.461*** |
| | (1.85) | (2.72) |
| Institutional ownership | 0.577*** | 0.567*** |
| | (5.86) | (5.76) |
| Family-firm × Inst. ownership | −0.802 | −1.659** |
| | (−1.42) | (−2.28) |
| Controls | Yes | Yes |
| Industry FE | Yes | Yes |
| Year FE | Yes | Yes |
| Observations | 8,592 | 8,592 |
| Pseudo $R^2$ | 0.206 | 0.206 |

*Notes.* This table reports PPML estimates examining whether institutional ownership moderates the relation between governance-based family-control constructs and CTM episodes. The dependent variable is the annual count of CTM episodes for each firm. All specifications include the same control variables as in Table 3, as well as industry and year fixed effects. Standard errors are clustered at the firm level. Parentheses report t-statistics. ***, **, and * indicate statistical significance at the 1%, 5%, and 10% levels, respectively.